\documentclass[osajnl,twocolumn,showpacs,superscriptaddress,10pt]{revtex4-1}
\usepackage{amsmath,amssymb,graphicx}

\DeclareMathOperator{\arccosh}{arccosh}
\usepackage{color}

\begin{document}

\title{Optical resonators based on Bloch surface waves}

\author{Matteo Menotti}
\affiliation{Department of Physics, University of Pavia, Via Bassi 6, I-27100 Pavia, Italy}
\email{matteo.menotti01@ateneopv.it}

\author{Marco Liscidini}
\affiliation{Department of Physics, University of Pavia, Via Bassi 6, I-27100 Pavia, Italy}

\begin{abstract}
A few recent works suggest the possibility of controlling light propagation at the interface of periodic multilayers supporting Bloch surface waves (BSWs), but optical resonators based on BSWs are yet to demonstrate. Here we discuss the feasibility of exploiting guided BSWs in a ring resonator configuration. In particular, we investigate the main issues related to the design of these structures, and we discuss about their limitations in terms of quality factors and dimensions. We believe these results might be useful for the development of a complete BSW-based platform for application ranging from optical sensing to the study of the light-matter interaction in micro and nano structures. 
\end{abstract}


\maketitle

\section{Introduction}\label{Sec:Introduction}
Light confinement near the surface of a photonic structure is crucial in a number of applications involving the light-matter interaction. For instance, it is appealing for the realization of optical sensors, in which one relies on the interaction between the electromagnetic field and the analytes under investigation. In this respect, surface plasmon polaritons are probably the most studied and utilized. Here, the interaction between the light and the free charges of the metal is exploited to achieve light confinement at a metallic surface and thus locally increase the electromagnetic field intensity \cite{barnes06}. This effect is used, for example, to increase the intensity of the Raman signal of a molecule or the efficiency of a fluorescent marker bonded to a target molecule \cite{prasad,homola03}. Yet, light confinement near the surface can also be exploited to control its propagation in the plane through a nano-structurization of the surface, which allows one to create waveguides and resonators with minimal fabrication requirements \cite{bozhevolnyi06}.

A promising strategy for such control and enhancement of light is based on photonic crystal (PhC) structures supporting Bloch surface waves (BSWs). These are evanescent electromagnetic field modes that propagate along the interface between a periodic dielectric stack and an homogeneous medium \cite{joannopulos,yeh78}. Their confinement relies on total internal reflection from the homogeneous medium and reflection within a photonic gap from the multilayer. Although more complicated than metallic systems, PhC-based devices are highly customizable, and they do not suffer from the absorption losses that plague metallic structures \cite{robertson05,guillerman07,liscidini07,liscidini09,delfan12}. 

High optical quality multilayers can be grown by molecular beam epitaxy, but they are also commercially available from thin-film companies for many materials, from semiconductor to oxides. More recently, the general features of guided BSW have been characterized theoretically and demonstrated experimentally \cite{Descrovi:1,Liscidini:1,Herzig:1}. All these results seem to suggest the opportunity for a generation of optical devices based on BSWs. This could be a very powerful approach for the development of integrated optical sensors, but also a new and versatile platform to probe interaction between photonic modes and electronic nanostructures in classical and quantum optics, for instance in fundamental studies of the light-matter interaction in the strong-coupling regime \cite{liscidini11,pirotta14,lerario14,yu14,angelini14}. All of this requires the demonstration of optical resonators, which are fundamental building blocks of any photonic platform, but are still lacking for BSWs structures.

In this work, we investigate the feasibility of realizing a ring resonator on the surface of periodic dielectric stacks supporting BSWs. In this structure the vertical confinement is provided by exploiting the surface state, and the lateral confinement is defined by a nano-fabrication of the surface, which could be done using either standard lithographic approaches or  low-cost imprinting techniques.  In Ref. \cite{Liscidini:1} Liscidini showed that this structure is extremely flexible and can support a variety of guided modes depending on the structure parameters and constituent materials.  Very recently, Wu \emph{et al.} reported on the fabrication and characterisation of bent waveguides supporting BSWs, demonstrating the feasibility of this geometry for more complex structures \cite{Herzig:1}.

Here we shall consider $TiO_2$/$SiO_2$ multilayers and Polymethyl-methacrylate (PMMA) ridges to work in the visible spectrum, for which the problem of designing small and integrated resonators is typically more demanding than in the infrared wavelength range due to the lack of transparent materials that can guarantee a large refractive index contrast. In this case one can work by etching a PhC crystal cavity in a membrane, but the resulting structure is usually very fragile \cite{mccutcheon08,martiradonna08}. Alternatively,  there have been demonstrated Hydex or SiN high-quality ring resonators embedded in a silica matrix \cite{little04,gondarenko09}, but in this case the region in which one experiences the largest light-matter interaction is not directly accessible. 

While most of the theoretical tools necessary for the design of multilayers are already available,  the design  and optimization of BSW waveguides and ring resonators is challenging. On the one hand, as the multilayered structure can be characterized by numerous layers, even a hundred, whose thicknesses can be as small as a few tens of nanometer, FDTD approaches can be largely time consuming and difficult to implement when the simulation cell is particularly large and high spatial resolution is required, as in this above-mentioned case. On the other hand, the typical size of the simulation in the plane of the multilayer, several micrometers, makes it difficult to describe the structure also in the reciprocal space by using Fourier Modal methods. Along the years, several strategies to solve for the confined modes in bent waveguides have been proposed: based on a  numerical solution of Maxwell equations in terms of an eigenvalue problem \cite{Marcatili:1}, expansion in Hankel wavefunctions in the cladding \cite{Marcuse:1}, conformal mapping of the bent waveguide \cite{Heiblum:1}, perturbative approaches  \cite{Melloni:1}, beam propagation methods \cite{Rivera:1}, finite element discretization \cite{Yamamoto:1}, etc...

Here, the different mechanisms at the base of the confinement of light in the vertical direction and in the plane of the structure suggest to follow a strategy based mainly on effective index approaches, which reduce the dimensionality of the problem and thus are able to quickly explore several configurations in the parameter space. In particular, our goal is to understand what are the main parameters that limit the quality factor achievable in an ideal BSW-based ring resonator when scattering losses due to fabrication imperfection can be neglected. Beside demonstrating the feasibility of a ring resonator approach, these results will serve as a guide in the design and development of a complete BSW-based platform.

The paper is structured as follows: in Sec. \ref{Sec:Structure} we describe the structure under investigation and the approach to the calculation of the mode quality factor of a PhC ridge ring resonator; in Sec. \ref{Sec:PerpendicularQ} we focus on the analysis of the intrinsic losses of BSW waveguide modes supported by a PhC ridge on a finite multilayer; in Sec. \ref{Sec:BendingQ} we deal with the bending losses associated with the typical lateral confinement that can be obtained for modes in PhC ridges; in Sec. \ref{Sec:DesignStrategy} we outline a general strategy to design BSW-based ring resonators. Finally, in Sec. \ref{Sec:Conclusions} we draw our conclusions.

\section{Structure and theoretical approach}\label{Sec:Structure}

\begin{figure}
    \centerline{\includegraphics[width=0.4\textwidth]{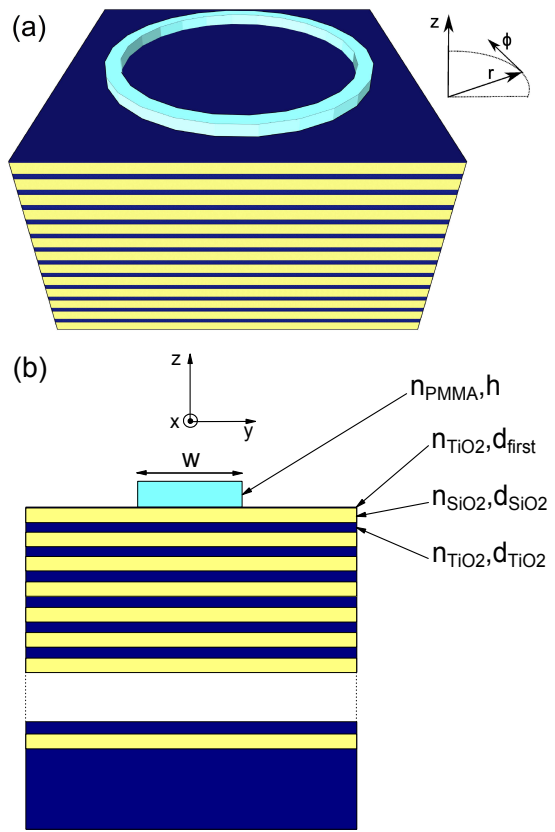}}
    \caption{(Color online) (a) Sketch of the PhC ridge ring resonator. (b) Cross section of the corresponding PhC ridge.}
    \label{Fig:Section}
\end{figure}

The structure we have in mind is a ring resonator of radius $R$ fabricated on the top of a multilayer supporting a BSW (see Fig. \ref{Fig:Section}(a)). The light confinement  in the vertical direction is due to the surface mode, while the lateral confinement is given by the PhC ridge, whose section is shown in Fig. \ref{Fig:Section}(b). This device could be fabricated using different kinds of dielectric materials, from semiconductors to oxides, in principle to operate at a given working point in the whole spectral region, with a bandwidth strongly dependent on the geometrical parameters and material choice. The structure we have in mind is a good compromise between micro-disk resonators \cite{Soltani:1}, which are typically realized in high-refractive index contrast platforms, and micro-pillar resonators \cite{gerard07}, which are less demanding in terms of refractive index contrast, but they are also more difficult to fabricate. The features of our structure are expected to be intermediate with respect to the above-mentioned systems: quality factors larger than those typically observed in micro-pillars (but smaller than what observed in micro-disks) along with flexibility in terms of fabrication and materials. In the following we will focus on a specific example to provide a description of the main features of this resonator.

The multilayer is periodic, with the unit cell composed of  two layers with thicknesses $d_\mathrm{TiO_2}=0.085$ $\mu$m and $d_\mathrm{SiO_2}=0.128$ $\mu$m, respectively. The multilayer has a finite number $N$ of periods and is truncated with the first layer made of $TiO_2$ with thickness $d_\mathrm{first}=0.010\ \mu m$. The structure is designed to operate in the visible spectral range at $\lambda_0=630$ nm (about $1.97$ eV) with $n_\mathrm{TiO_2}=2.58534$ and $n_\mathrm{SiO_2}=1.54270$. The choice of these materials for the multilayer is  convenient from a technological point of view, for they are commonly employed in the fabrication of optical filters by sputtering, thus this plan structure is commercially available. Finally, we consider a PMMA ($n_\mathrm{PMMA}=1.48914$) ridge with height $h=0.220$ $\mu m$ and width $w=0.800$ $\mu m$ on the top of the multilayer.

\begin{figure}
   \centerline{\includegraphics[width=0.47\textwidth]{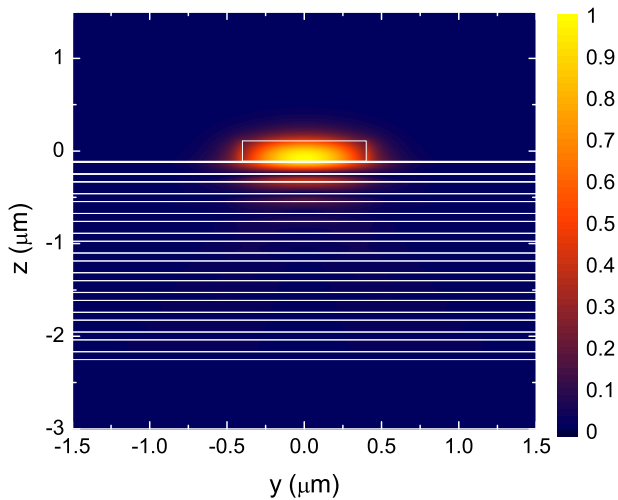}}
    \caption{(Color online) Intensity profile in the $yz$-plane of a TE-polarized guided BSW calculated at $\lambda_0 = 0.630\ \mu m$. The PhC is made of $N=10$ periods of a $TiO_2$/$SiO_2$ unit cell ($d_\mathrm{TiO_2} = 0.085\ \mu m$ and $d_\mathrm{SiO_2} = 0.128\ \mu m$) and the first $TiO_2$ layer is truncated with $d_\mathrm{first} = 0.010\ \mu m$. The ridge material is PMMA with a rectangular section ($0.800\ \mu m\times0.220\ \mu m$). The profile is obtained by means of a FDTD numerical simulation.}
    \label{Fig:Profile}
\end{figure}

The structure shown in Fig. \ref{Fig:Section}(b) is symmetric upon reflection with respect to the $xz$-plane. Consequently, the guided modes can be classified according to the eigenvalues of the mirror operator $\hat{\sigma}_{xz}$ in TE (Transverse-Electric, $\hat{\sigma}_{xz}=-1$) and TM (Transverse-Magnetic, $\hat{\sigma}_{xz}=+1$) modes. In the following we shall focus only on the fundamental TE mode, whose intensity profile, calculated by FDTD \cite{Taflove:1,Lumerical:1}, is plotted in Fig. \ref{Fig:Profile}. This mode is a perturbation of the BSW supported by the bare multilayer, and light is guided within the ridge likewise in a rib waveguide. It should be noticed that, unlike truly guided BSW, this mode can exist for any choice of the ridge height, with its properties and modal volume depending strongly on $h$ \cite{Liscidini:1}. Here the ridge height has been chosen to guarantee a good lateral confinement and, at the same time, keep the modal volume as small as possible.

One can obtain a ring resonator by bending the straight waveguide onto itself in a circular shape (see Fig. \ref{Fig:Section}(a)). The resonant frequencies are obtained by the usual condition \cite{Yariv:1}
\begin{equation}\label{resonant_condition}
kL\equiv\frac{2\pi}{\lambda_0}n_\mathrm{mode}2\pi R=2\pi m,
\end{equation}
where $n_\mathrm{mode}$ is the real part of the effective index of the mode in the bent waveguide, $R$ is the ring radius, and $m$ is an integer.
In the absence of the $xz$-plane mirror symmetry, a rigorous TE/TM classification of the guided modes is no longer possible. However, in waveguides with a rectangular cross section, the modes having the $E_r$ component dominant over $E_{\phi}$ and $E_z$ \cite{Shih:1} can be referred to as \emph{TE-like}, and in the following we shall always consider these class of solutions. Finally, this structure is characterized by a continuous rotational symmetry and therefore, by choosing cylindrical coordinates $(r,\phi,z)$, the angular dependence of the fields can be written in the form $e^{im\phi}$, with $m$ integer. This property turns out to be extremely useful in the numerical description of the structure, as it allows one to reduce the dimensionality of the electromagnetic problem, for example in FDTD calculations.

Our goal is the design of a ring resonator working at $\lambda_0$ and the computation of the mode quality factor, which can be used to measure the performances of the device. In the case of a BSW-based ring resonator, this task is all but simple, as an analytical approach is not feasible, and even a \emph{brute force} 3D numerical calculation, for example by using  a FDTD method, can be extremely time-consuming on a standard personal computer or medium-sized server. Thus, in the view of optimizing the structure, which depends on a large number of parameters (unit cell composition, multilayer truncation, height and width of the ridge, ring radius, etc...), one should rather look for an approximated but faster strategy. 

The structure and the mode profile in Fig. \ref{Fig:Profile} seem to suggest a possible path towards the solution of the problem. Indeed, the strong field confinement of light in the ridge waveguide indicates that the resonant modes properties can be found by decomposing the 3D quest in two independent 1D and 2D problems. First, we describe the field confinement due to the surface state in an effective multilayer, then we find the resonant modes by modelling an effective 2D system using a FDTD method, which allows us to take into account also for the dependence of $n_\mathrm{mode}$ on the ring radius $R$. Similar approaches have been used to study lower-order  \cite{gerard96} and whispering gallery modes \cite{gerard07} in micro-pillar resonators, demonstrating surprising effectiveness even for $R$ of a few $\mu$m. Very recently, the same strategy has been successfully  applied to describe light confinement and propagation of BSWs in bent waveguides \cite{Herzig:1}.

It should be noticed that the mode quality factor $Q$ depends on the light leakage through the substrate, due to the finite number $N$ of periods in the multilayer, and the bending losses, which, for a given ridge, are a function of the ring radius $R$. One can start by considering that $Q$ measures the energy loss experienced by the electromagnetic mode per cycle and, assuming that the two loss mechanisms are independent, can write
 \begin{equation}
 \frac{1}{Q}=\frac{1}{Q_{\perp}}+\frac{1}{Q_{\parallel}},
\end{equation}
where $Q_{\perp}$ is the quality factor determined by the out-of-plane losses and $Q_{\parallel}$ is the quality factor limited by the in-plane losses. Thus, first we deal with the description of light leakage in the substrate, then we compute $Q_{\parallel}$. It should be noticed that here we are neglecting extrinsic losses, such as scattering at the interface between layers due to surface roughness or fabrication imperfection. In a real device these contributions can be important, depending on the accuracy in the fabrication processes.

\section{Multilayer design and  $Q_{\perp}$}\label{Sec:PerpendicularQ}

Here we consider a sufficiently large ring radius $R$ so that the mode field profile and propagation wavevector are slightly affected by the curvature, i.e. the effective index $n_\mathrm{mode}$ in \eqref{resonant_condition} is close to that of a straight waveguide (see also Sec. \ref{Sec:BendingQ}), and the vertical and horizontal dynamic can be decoupled. For this reason, the quality factor $Q_{\perp}$ can be computed by determining the propagation losses of the corresponding guided BSW (GBSW) in a straight ridge waveguide having the cross-section shown in Fig. \ref{Fig:Section}(b). In particular, if one assumes a semi-infinite PhC, the mode is truly guided as there are no propagation losses. On the contrary, in a more realistic situation, when  the multilayer has a finite number $N$ of periods, the GBSW is characterized by a finite propagation length, even when scattering losses can be neglected, as light can tunnel trough the multilayer into the substrate. 

The analysis of light confinement in the $yz$ plane for a straight waveguide could be performed by using FDTD calculations or other 2D numerical approaches. Yet, in the spirit of suggesting a reliable but also fast approach that would allow one to rapidly explore the parameter space of the structure, we adopt an effective index method (EIM) by separating the field confinement in the $y$ and $z$ direction. It should be noticed that this approximation has already been proved to be reliable to study analogous structures \cite{Herzig:1,Bienstman:1, Bienstman:2, Eguchi:1, Qiu:1}. More recently, the accuracy of this strategy for the description of guided modes in PhC ridges has been investigated in Ref. \cite{Liscidini:1}, but the issue of propagation losses has not been discussed.

The EIM is conveniently employed to decouple the field confinement in 2D electromagnetic problems, provided that light is well confined in the ridge at least in one dimension. Thus a 2D problem is reduced to the composition of two 1D tasks, easily solved by means of the transfer matrix method (TMM).

First, the EIM fictitiously collapses the guiding 2D structure in the direction where the most intense light confinement takes place. This is obtained by dividing the structure in 1D multilayers (in accordance with the guiding regions shape) and calculating in each region the confined mode profiles and their effective indices. Subsequently, the structure in each slice is replaced by an homogeneous layer with a refractive index $n_\mathrm{eff}$ corresponding to that of a proper confined mode.
Therefore, one can treat the most significant light confinement mechanism in each section exactly, and obtain an effective 1D structure along the complementary direction, where the less intense localization of the field takes place. This allows the computation of the 2D field profile, the mode dispersion relation, and the propagation losses.

One should note that the EIM predictions, although approximated, are generally very close to the  values calculated by more time-consuming 2D numerical methods. However, the goodness of this approach depends on the structure geometry and field distribution.

Referring to Fig. \ref{Fig:Profile}, it is clear that in our case the field confinement is strong in both $y$ and $z$ direction. For this reason, we expect that dividing the structure in the vertical or in the horizontal direction (although this implies different approximations) should produce quite similar and accurate results.
In the following we will study the properties of the confined modes supported by our structure employing both the strategies.

\subsection{Horizontal EIM}
\begin{figure}
    \centerline{\includegraphics[width=0.5\textwidth]{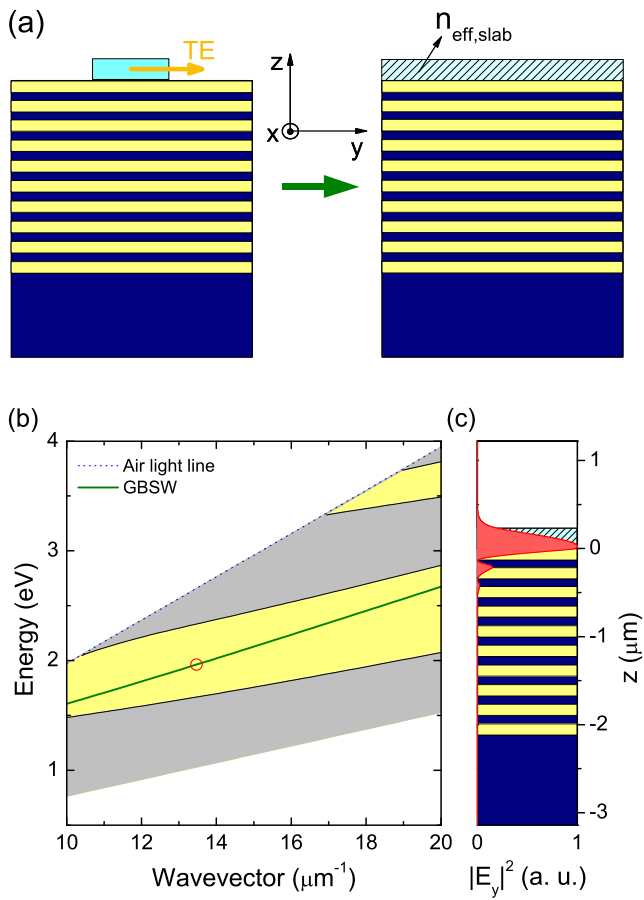}}
    \caption{(Color online) (a) Sketch of the dimensionality reduction of the electromagnetic problem by using the HEIM. (b) Real part of the GBSW dispersion relation. The yellow-shaded regions represent the PBGs. The dispersion relation is calculated considering the modal dispersion in the first effective layer. (c) $E_{y}$ field intensity in the truncated multilayer.}
    \label{Fig:Dispersion}
\end{figure}

Here we exploit the strength of the vertical confinement of the field, due to the PBG mechanism on the PhC side and the TIR phenomenon at the PMMA/Air interface, and hence we divide the structure along the $z$ direction.
Following the EIM described above, one substitutes the PMMA ridge with an effective homogeneous layer of thickness $h$ and refractive index $n_\mathrm{eff,slab}(\omega)$, corresponding to that of the fundamental TM guided mode supported by a symmetric PMMA slab waveguide of width $w$ in air. Note that here TM refers to the plane of the PMMA slab waveguide. This is equivalent to the horizontal cross section of the PMMA ridge (see Fig. \ref{Fig:Dispersion}(a)), thus we shall refer to this method as the \emph{horizontal} EIM (HEIM). The effective index can be written as $n_\mathrm{eff,slab}=\kappa_\mathrm{eff,slab}c/\omega$, where $\kappa_\mathrm{eff,slab}$ is the zero of the $T_{22,\mathrm{slab}}(k,\omega)$ element of the slab transfer matrix. Finally, one can calculate the dispersion relation of the GBSW by finding the guided modes of this effective multilayer, searching for the solutions of the equation $T_{22,\mathrm{mul}}(k,\omega)=0$. In general, these are solutions of the kind $\kappa_\mathrm{GSBW}=\beta+i\gamma$, where $\beta=(\omega/c)n_\mathrm{GBSW}$ and $\gamma=1/(2L_\mathrm{prop})$, with $L_\mathrm{prop}$ the mode propagation length.

In  Fig. \ref{Fig:Dispersion}(b) we show the real part of the dispersion curve corresponding to our parameters.
As expected, the mode dispersion relation is within the PBG and below the air light line. In particular, for $\lambda_0=630\ nm$ we have $\beta=13.5164\ \mu m^{-1}$ and $n_\mathrm{GBSW}=1.3553$; the corresponding mode intensity profile is plotted in Fig. \ref{Fig:Dispersion}(c).

For a sufficiently large $N$ (typically more than $5$, depending on the refractive index contrast of the materials), the dispersion relation is independent of the number of periods of the multilayer. On the contrary, the mode propagation length depends exponentially on $N$, for $L_\mathrm{prop}$ is directly related to the imaginary part  $\alpha$ of  the Bloch wavevector, which controls the exponentially damping of the electric field in the multilayer.  In general, one has 
\begin{equation}\label{Eq:Lprop}
  L_\mathrm{prop}\propto\frac{1}{2}\exp[2N\alpha\Lambda],
\end{equation}
where $\alpha$ is
\begin{equation}
  \alpha=\frac{1}{\Lambda}\arccosh\left(\frac{Tr[T_{\Lambda}]}{2}\right),
\end{equation}
in which one has applied the Bloch-Floquet theorem to a generic PhC with period $\Lambda$ and transfer matrix $T_{\Lambda}$ \cite{Yariv:1}.

\begin{figure}
   \centerline{\includegraphics[width=0.47\textwidth]{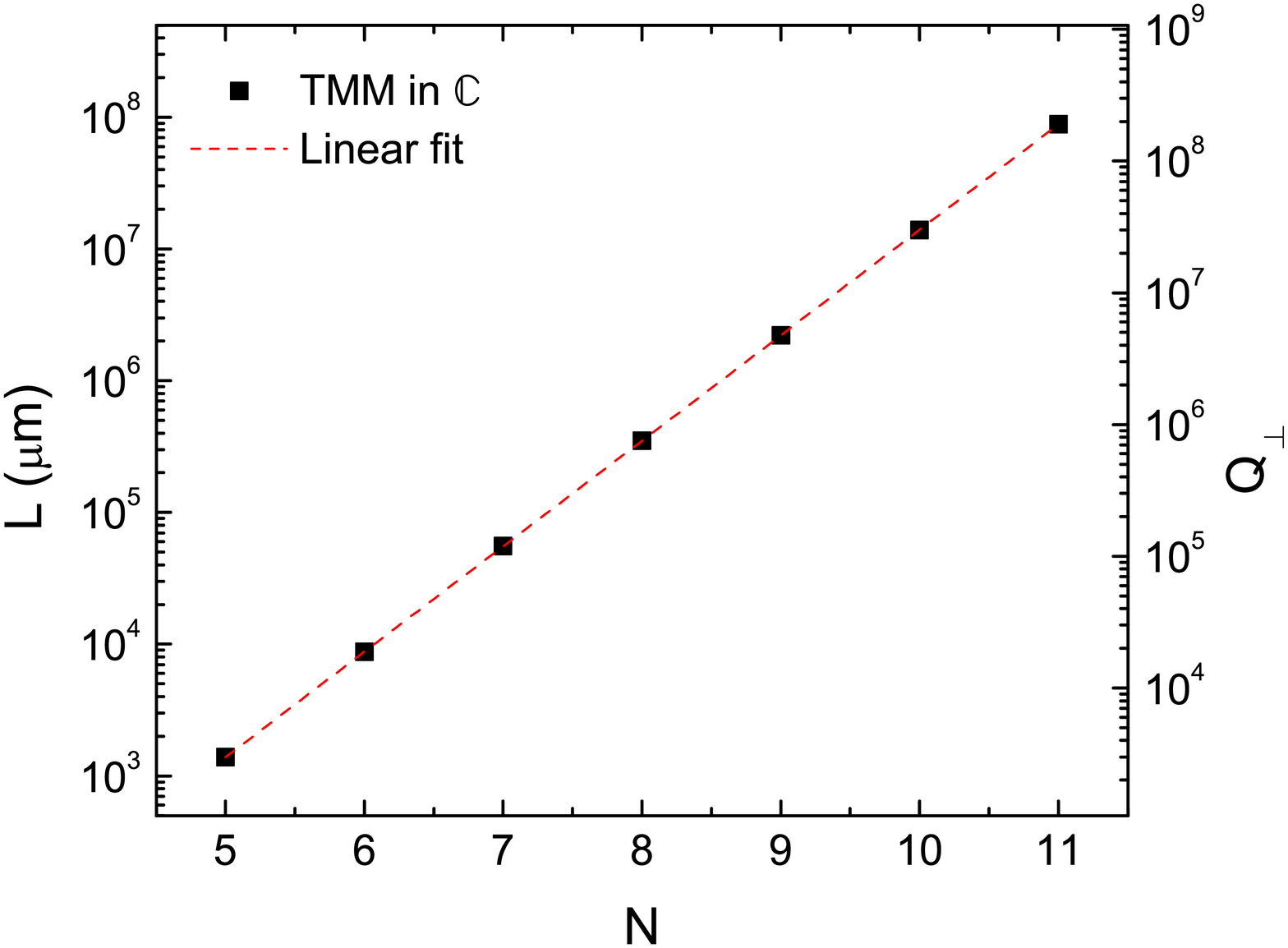}}
    \caption{(Color online) Propagation length $L$ as a function of the number of periods $N$ in the multilayer, computed by means of the HEIM. On the right axis the corresponding perpendicular quality factor $Q_{\perp}$ (when Fig. \ref{Fig:Section}(b) is the section of a ring resonator) is reported.}
    \label{Fig:QvsN_HEIM}
\end{figure}

Since the quality factor of the resonator can also be viewed as the number of field oscillations in the ring before the field intensity is reduced by a factor $1/e$, we can also write $Q_{\perp}=L_\mathrm{prop}/\lambda=\beta/(4\pi\gamma)$. In Fig. \ref{Fig:QvsN_HEIM} we show $Q_{\perp}$ calculated from the GBSW complex wave vector. As expected, we have an exponential growing of $Q_{\perp}$ with $N$. It should be noticed that in our structure one can obtain propagation distances of about $10\ m$ with only $10$ periods. In particular, this increasing is well fitted by a curve of slope ($2\alpha\Lambda$) as suggested by Eq. \eqref{Eq:Lprop}. Thus, our results show that a fast and reliable optimization of the multilayer can be done very easily by computing the imaginary part of the Bloch wave vector as a function of the sole unit cell composition.

\begin{figure}
  \centerline{\includegraphics[width=0.47\textwidth]{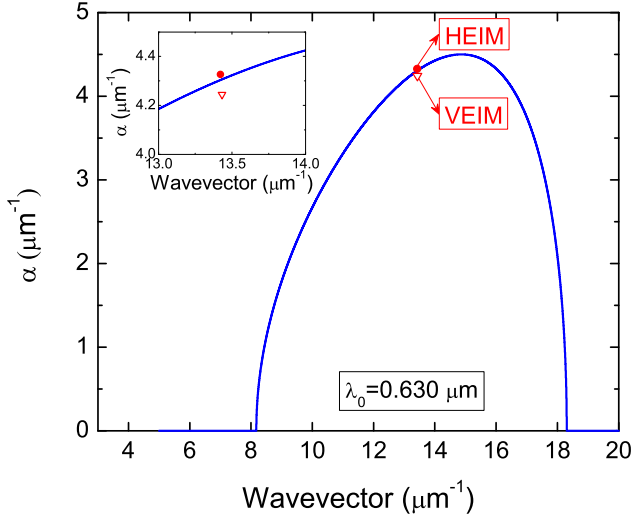}}
    \caption{(Color online) Imaginary part of the Bloch wavevector in a PhC with $TiO_2/SiO_2$ unit cell ($d_\mathrm{TiO_2}=0.085\ \mu m$, $d_\mathrm{SiO_2}=0.128\ \mu m$), as a function of the wavevector. The curve is calculated at $\lambda_0=0.630 \ \mu m$. The working point corresponding to our choice of the multilayer truncation and the ridge geometry, calculated both with the HEIM and the VEIM, is reported.}
    \label{Fig:Imaginary}
\end{figure}

In Fig. \ref{Fig:Imaginary} we plot the imaginary part of the Bloch wavevector as a function of $\beta$ at $\lambda_0=630$ nm for our structure parameters.
We also indicate the working point corresponding to our final choice of the multilayer truncation and the ridge parameters. It should be noticed that this point does not coincide with the curve maximum, which would guarantee the fastest field decay in the multilayer. Indeed, although one usually operates in the region of large $\alpha$, the BSW ring resonator is designed to minimize also the bending losses in the ring, which of course depend on the ridge parameters. Thus, in general, the optimal parameters do not necessarily lead to the fastest growing of $Q_{\perp}$ with $N$, due to a compromise between vertical and lateral light confinement.

\subsection{Vertical EIM}
\begin{figure}
	\centerline{\includegraphics[width=0.5\textwidth]{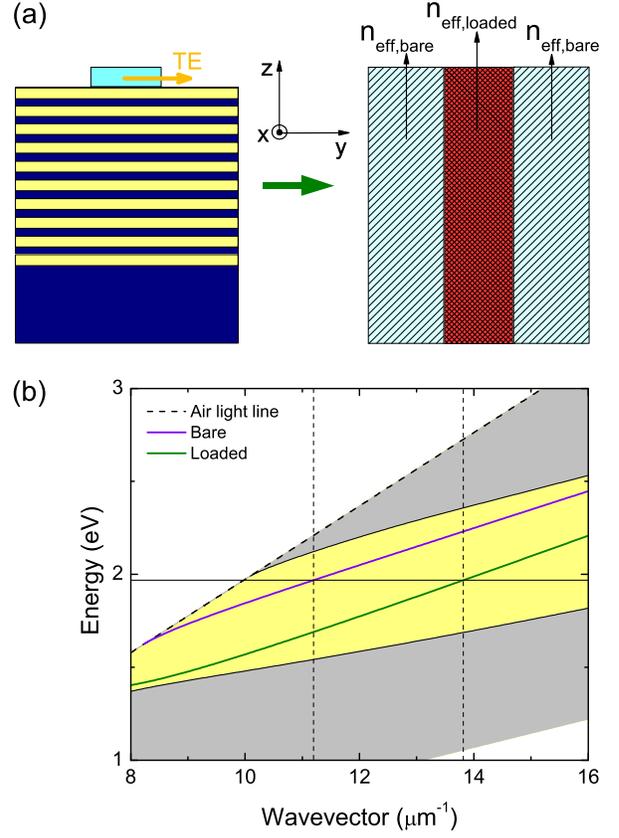}}
    \caption{(Color online) (a) Sketch of the dimensionality reduction of the electromagnetic problem by using the HEIM. (b) Real part of the dispersion curves for the bare and loaded modes. The yellow-shaded region represents the PBG.}
    \label{Fig:Dispersion_VEIM}
\end{figure}

The field intensity profile reported in Fig. \ref{Fig:Profile} shows that a strong confinement of the GBSW takes place also in the $y$ direction, due to the TIR mechanism at the PMMA/Air interfaces.
Thus, one can obtain results similar to those shown in the previous section following the EIM prescription and dividing the structure vertically in two different regions, named \emph{bare}, where the multilayer cladding is simply air, and  \emph{loaded}, in which the multilayer has an additional layer of height $h$ corresponding to ridge region (see Fig. \ref{Fig:Dispersion_VEIM}(a)). We search for the TE surface states supported by each multilayer, and we calculate the corresponding effective indices $n_\mathrm{eff,bare}$ and $n_\mathrm{eff,loaded}$ at $\lambda_0$. The straight waveguide is finally approximated by a vertical slab of index $n_\mathrm{eff,loaded}$ with cladding of index $n_\mathrm{eff,bare}$. Finally, one searches for the fundamental TM mode, whose wavevector is $\kappa_\mathrm{GBSW}=\beta+i\gamma$. Note that here TM refers to the plane of the effective slab. We shall refer to this second effective index approach as the \emph{vertical EIM} (VEIM, see also Ref.\cite{Liscidini:1}).

The dispersion curves associated to the bare and loaded multilayers are reported in Fig. \ref{Fig:Dispersion_VEIM}(b), and the effective indices calculated at $\lambda_0=630\ nm$ for $N=10$ are $n_\mathrm{eff,bare}=1.1225+2.0602\cdot10^{-7}i$ and $n_\mathrm{eff,loaded}=1.3853+2.2347\cdot10^{-9}i$, respectively. The resulting GBSW wavevector is $\kappa_\mathrm{GBSW} = (13.4332 + 1.2205\cdot10^{-7})\ \mu m^{-1}$.

\begin{figure}
   \centerline{\includegraphics[width=0.47\textwidth]{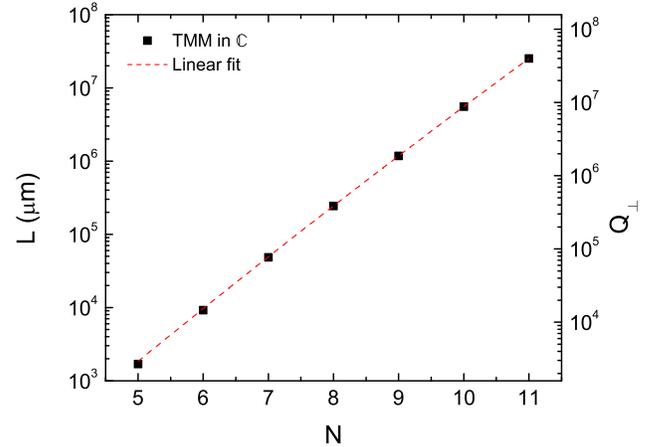}}
    \caption{(Color online) Propagation length $L$ as a function of the number of periods $N$ in the multilayer, computed by means of the VEIM. On the right axis the corresponding perpendicular quality factor $Q_{\perp}$ (when Fig. \ref{Fig:Section}(b) is the section of a ring resonator) is reported.}
    \label{Fig:QvsN_VEIM}
\end{figure}

In Fig. \ref{Fig:QvsN_VEIM} we show $Q_\perp$ calculated by means of the VEIM as a function of the number of periods in the PhC.
Similarly to the HEIM case, $Q_\perp$ increases exponentially with $N$, and the slope is very close to that shown in Fig. \ref{Fig:QvsN_HEIM}.
For a direct comparison, we report the working point obtained with the VEIM in Fig. \ref{Fig:Imaginary} along with the HEIM one. The two methods show a good agreement in the prediction of both the propagation constant $\beta$ and the losses of the GBSW.

\section{Ring design and  $Q_{\parallel}$  }\label{Sec:BendingQ}

\begin{figure}[htbp]
   \centerline{\includegraphics[width=0.47\textwidth]{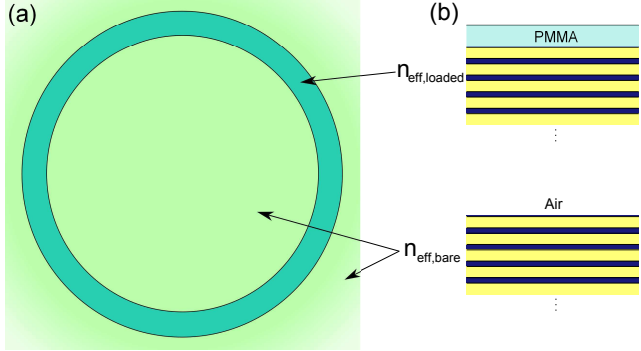}}
    \caption{(Color online) Top view of the circularly bent effective slab waveguide. The light blue region refractive index is $n_\mathrm{eff,loaded}=1.3853$ while the surrounding medium is characterized by $n_\mathrm{eff,bare}=1.1225$.}
    \label{Fig:Ring_Top}
\end{figure}

Here, the calculation of the in-plane quality factor $Q_\parallel$ is performed only by means of a combination of VEIM and FDTD simulation in cylindrical coordinates, which is analogous to the approach adopted in Ref. \cite{Herzig:1}, where it is experimentally validated. First, the structure is divided in bare and loaded multilayer regions (see Fig. \ref{Fig:Ring_Top}(b)), with refractive indices $n_\mathrm{eff,bare}$ and $n_\mathrm{eff,loaded}$, respectively. This allows us to approximate the 3D structure with a 2D ring resonator (see Fig. \ref{Fig:Ring_Top}(a)).

The computation of $Q_{\parallel}$ around $\lambda_0$ is performed using Meep, a free FDTD simulation software package developed at MIT \cite{Meep:1}. This code is characterised by the possibility of exploiting the continuous rotational symmetry of the ring by working in cylindrical coordinates and solving the angular dependence of the fields analytically, according to $e^{im\phi}$, with $m$ integer. This reduces the calculation to a simple 1D simulation, with large benefits in terms of the computational effort. 

\begin{figure}[htbp]
 \centerline{\includegraphics[width=0.47\textwidth]{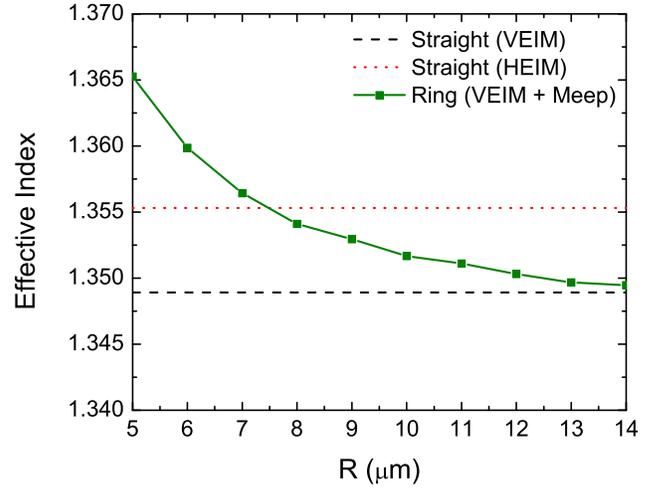}}
    \caption{(Color online) Effective index of the GBSW calculated by means of the VEIM  and 2D FDTD as a function of the ring radius $R$. The values calculated with the HEIM and the VEIM for the corresponding straight waveguide are also shown.}
    \label{Fig:neff_all}
\end{figure}

This approach takes into account the modification of the mode effective index due to the waveguide bending. It is interesting to compare the results calculated for the straight ridge by means of both the HEIM or VEIM with the value of the mode effective index in the case of a bent waveguide. The comparison is shown in Fig. \ref{Fig:neff_all}. As expected, for larger $R$, the effective FDTD results approach the mode index calculated with the VEIM. It should be noticed that in the range of $R$ considered in our work all the three values are very close, with differences corresponding to an error in the determination of the resonance frequency less than the free spectral range of the ring resonator. This confirms the goodness of our approximation for the parameters under consideration and highlights the possibility of determining the resonance frequency (but not the losses) by means of the analysis of the sole straight waveguide, even in the case of relatively small $R$.

\begin{figure}[htbp]
 \centerline{\includegraphics[width=0.47\textwidth]{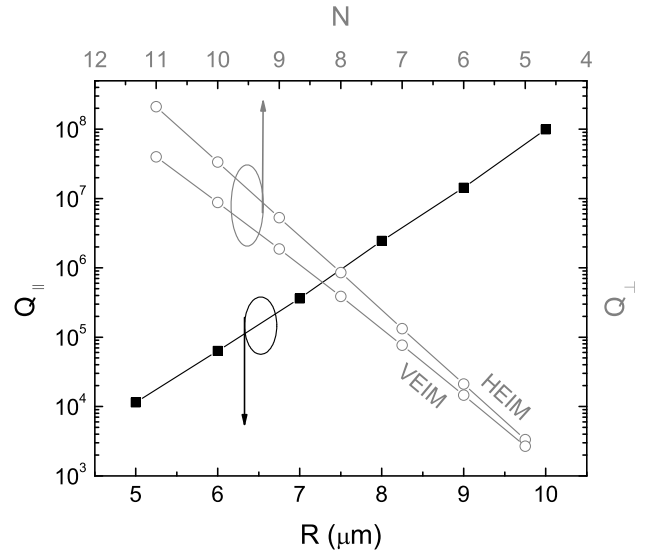}}
    \caption{(Color online) Lateral quality factor $Q_{\parallel}$ as a function of the bending radius $R$ calculated by means of the VEIM and FDTD numerical simulations. The result is compared with the perpendicular quality factors $Q_\perp$ obtained in section \ref{Sec:PerpendicularQ}.}
    \label{Fig:QvsR}
\end{figure}

In Fig. \ref{Fig:QvsR} we show $Q_{\parallel}$ as a function of the ring radius. As expected from a standard treatment of bending losses, $Q_{\parallel}$ increases about exponentially with $R$. We also show $Q_{\perp}$ versus $N$, so that one can immediately see the essential requirements in order to meet a given $Q$ target.

It should be noticed that the value of the quality factors in such a small ring resonator exceeds those that are experimentally observed in structures characterized by a stronger refractive index contrast \cite{Xu:1} or larger radius $R$ \cite{Kurz:1}.
Naturally, these theoretical values are obtained by neglecting scattering losses and fabrication imperfections, which reduce the quality factor dramatically.
However, the quality factors calculated here are so high, that even if scattering losses (and every other extrinsic loss mechanism) will affect the resonator considerably, the actual device would still display good light confinement properties.
For instance, let us consider a ring resonator with $N=10$ periods and radius $R=9\ \mu m$. For this structure the quality factor is $Q\sim10^7$ and even allowing the scattering losses to lower this value by three orders of magnitude, we would still obtain a ring resonator with $Q\sim10^4$ within a footprint of $\sim250\ \mu m^2$.

\section{Design strategy and final remarks}\label{Sec:DesignStrategy}
The design of a BSW-based ring resonator to achieve a target quality factor is critically dependent on the constituent materials, as well as on the PhC truncation, and on the ridge section. For this reason, it is useful to outline a basic strategy for designing and optimizing BSW-based resonators.

In the previous sections we have considered a specific $TiO_2$/$SiO_2$ BSW ring resonator working at $\lambda_0=630$ nm. Although the final structure depends critically on the constituent materials and target wavelength, it is possible to outline a general design strategy to obtain structures to work in a different spectral region and characterised by a target quality factor and  modal volume. The latter is essentially determined by the exponential  decay of the electromagnetic field in the multilayer and the ring radius. Despite the main goal of this work is not the optimization of a structure for a specific application, one can be guided in the design of new BSW ring resonators by a few simple rules.  

As we have seen, a BSW ring resonator is made up of several layers and different materials, and it is characterized by a rather complex geometry that makes the structure modelling quite challenging. Indeed, it requires to perform several numerical simulations (looped over all the variables) to cover the vast parameters space. In this regard, the EIM-based approach we have proposed here is actually a valuable tool, suitable for this purpose. It allows to obtain the relevant physical properties of the confined electromagnetic modes introducing minimal errors while requiring a very small computational cost. Yet, it should be noticed that, even in this case, one cannot explore such a large parameter space at the same time. 

Once the target wavelength is chosen, one starts by \emph{optimising the unit cell of the multilayer} that will support the entire structure. Given the  materials compatible with the most accessible fabrication technique and the specific application one has in mind, the best strategy is looking for strong refractive index contrast and transparency. Indeed, this will lead to a large PBG that offers flexibility in the design of the PhC ridge and resonator.  This will also determine high $Q_{\perp}$s with small N, for the larger is the index mismatch in the unit cell, the faster is the field exponential decay in the multilayer. For the choice of the layers' thickness, a good thumb-rule is considering quarter-wave stack, where this condition has to be satisfied at the working point below the light line corresponding to the upper cladding material and not at normal incidence as it is usually done in distributed Bragg reflectors. 

The second step is choosing the \emph{multilayer truncation and ridge parameters}, which determine the main properties of the mode supported by the PhC ridge. In this communication we have dealt with a mode characterised by a strong field confinement in the ridge region. However,  a PhC ridge can support a larger variety of modes, which all could be used for the realization of BSW ring resonators (see e.g. \cite{Liscidini:1}). 

With respect to the case investigated in this paper, the position of the bare and loaded modes inside the PBG at a given $\lambda_0$ is crucial in the optimisation of both $Q_\perp$ and $Q_\parallel$. On the one hand, the best vertical confinement of the surface state is obtained when both modes are close to the PBG center (where $\alpha$ is highest). On the other hand, one looks also for a large contrast $\Delta n_\mathrm{eff}=n_\mathrm{eff,loaded}-n_\mathrm{eff,bare}$, which guarantees better lateral confinement and small bending radii. These two conditions point in opposite directions in the parameter space, thus one looks for a \emph{compromise between lateral and vertical confinement}. In our case, without any particular application in mind, we have designed our structure to maximize the effective index mismatch by making the bare mode effective index close to the air light line; the loaded effective index is engineered to be as close as possible to the PMMA light line while maintaining at the same time the vertical waveguide single mode. This results in a very compact BSW ring resonator, characterised by large quality factors. 

The last step is the \emph{structure trimming} by using a more accurate fully-3D simulation, for example by means of FDTD. Yet, depending on the size of the final structure, this might be hard, if not impossible, on a standard personal computer. 

\section{Conclusions}\label{Sec:Conclusions}
In this work we proposed an original resonator consisting in a photonic crystal ridge supporting a guided BSW and bent in a circular shape. We have studied its optical properties, including the dependence of the quality factor on both the ring size and multilayer composition. We outlined a very general strategy that could be used to design and optimize BSW ring resonators made of different materials and operating in different spectral regions. This approach is based on assuming vertical and in-plane optical losses as independent, and computing the corresponding perpendicular quality factor  $Q_\perp$ and in-plane quality factor $Q_\parallel$ by means of  TMM and effective 2D FDTD calculations. As an example,  we designed and analyzed a  BSW ring resonator constituted of a PMMA ridge waveguide on a truncated periodic $TiO_2/SiO_2$ multilayer, operating  at $\lambda_0\approx 0.630$ $\mu$m, having a theoretical value of $Q$ exceeding  $10^7$, and a device footprint of $\sim250$ $\mu$m$^2$.

Given the strong interest in BSWs for the realization of integrated optical sensors and as a tool to investigate the light-matter interaction at the fundamental level, and the recent experimental demonstration of guided BSWs in bent waveguides, we believe these results might be of great interest to that part of the community working in integrated optics.

\end{document}